\documentclass[global,twocolumn]{svjour}

\usepackage{latexsym}
\usepackage{graphics}
\usepackage{graphicx}
\usepackage{amsmath}
\begin{document}
\title{Dynamic Terahertz Beam Steering Based on Graphene Metasurfaces}
\author{Liming Liu\inst{1,2}\thanks{\emph{email:} liming.liu@student.adfa.edu.au}, Yair Zarate\inst{2}, Haroldo T. Hattori\inst{1}}
%
%
%
\institute{School of Engineering and Information Technology, University of New South Wales, ACT 2612, Australia. \and Nonlinear Physics Centre, Research School of Physics and Engineering, Australian National University, Canberra, Australia.}
%
\date{December 17, 2015}

%
\maketitle
\begin{abstract}
A full (2$\pi$) phase modulation is critical for efficient wavefront manipulation. In this article, a metasurface based on graphene long/short-strip resonators is used to implement a dynamic 2$\pi$ phase modulation by applying different voltages to different graphene resonators. The configuration is found to have high reflection efficiency (minimum 56\%) and has a full phase modulation in a wide frequency range. Terahertz (THz) beam steering as large as 120 degrees ($\pm60^\circ$) is demonstrated in a broad frequency range (1.2 to 1.9 THz) by changing the Fermi levels of different graphene resonators accordingly. This metasurface can provide a new platform for effectively manipulating THz waves.
\end{abstract}
\section{Introduction}
\label{intro}

The past decade has witnessed the advent of metamaterials that are capable of manipulating electromagnetic waves in an unprecedented way-for example, by controlling permittivity and permeability of materials. Exotic properties have been demonstrated including negative refractive index~\cite{smith}, super lens~\cite{Pendry} and cloaking~\cite{Cloaking}. More recently, metasurfaces, two-dimensional (2D) counterparts of metamaterials, have shown the ability of fully controlling the polarization, amplitude and phase of electromagnetic waves~\cite{Zhao2011,Shulin,reviewer1,reviewer2,reviewer3}. Distinct from thick optical components, such as lenses, which rely solely on phase accumulations along the optical path, metasurfaces utilize the abrupt phase shift at the interface produced by localized resonances. With a full phase modulation of electromagnetic waves, arbitrary wavefront manipulations can be achieved and new functionalities have been demonstrated, including flat lens~\cite{Quanlong}, anomalous reflection or refraction~\cite{Nanfang} and hologram~\cite{Hologram}. 

A full (2$\pi$) phase modulation is usually required for efficient wavefront control and it is desirable to have a constant amplitude at the same time. In general, tunable resonant metamaterials are utilized to produce a continuous full phase modulation. Common ways to create tunable metamaterials include changing the geometries of metamaterials, modifying their surrounding media, and manipulating the coupling between adjacent metamaterials. In particular, numerous works on tunable metamaterials have been carried out at the THz range where natural materials do not interact strongly with THz waves~\cite{Willie}. For instance, Microelectromechanical (MEMS) based tunable THz metamaterials have demonstrated large tunability and the potential to produce highly integrated THz devices~\cite{Ma2014MEMS}. Tunability in liquid crystals based metamaterials have also been demonstrated by applying an external voltage or changing the temperature of the device~\cite{Liu2013,Willie2013}. Manipulating the coupling between adjacent metamaterial layers are experimentally shown to be an effective way to achieve large tunability in the THz regime~\cite{Liu2014}. However, those tunable metamaterials only focus on frequency tunability and cannot achieve large phase modulation.

To achieve a 2$\pi$ phase modulation, three different approaches are primarily used. The first approach relies on cross-polarization to impose a 2$\pi$ phase shift in the transmitted waves, which usually lead to a low working efficiency~\cite{V-shape}. The second way is to use dielectric Huygens’ surfaces by overlapping electric and magnetic resonances to produce a 2$\pi$ phase shift with high efficiency~\cite{Decker}. Another approach is to use a metallic ground plane and one resonator separated by a dielectric spacer~\cite{Bozhevolnyi}. Due to the intrinsic Lorentz-like resonance response, the phase shift produced by one resonance is limited to $\pi$ radians. By including a metallic ground plane an additional phase is added to cover the full 2$\pi$ range in the vicinity of the resonance. By working in reflection, higher efficiency is generally expected.

In real-world applications, it is crucial to implement metasurfaces with a dynamic phase tunability. However, so far, most metasurfaces are static since they tune the whole metasurface uniformly~\cite{tunable_metasurface}, which eliminates the advantage of having gradient metasurfaces. The challenge of having a dynamic tunability for metasurfaces mainly comes from fabrication difficulties to control each discrete resonator separately.

The emerging 2D material graphene is suitable for building 2D metasurfaces with a dynamic phase modulation. Particularly, graphene behaves like a metal in the THz range and its conductivity can be tuned by applying an external voltage~\cite{Graphene}. This special property has generated a lot of interest to build graphene based tunable metamaterials~\cite{Ilya-graphene,Weiren,Graphene_THz1}. In the previous work, we have proposed a metasurface based on double graphene resonator in one unit cell, which can effectively focus different THz wavelengths into the same spot~\cite{new_apl}. In this article, we further discuss more details on the configuration and a THz beam steering is demonstrated with steering angle as large as 120$^{\circ}$ from 1.2 to 1.9 THz with a minimum of 56\% reflection efficiency. 

\section{Graphene Metasurfaces}

\begin{figure}[tb]
\centering
\protect\includegraphics[width=0.9\columnwidth]{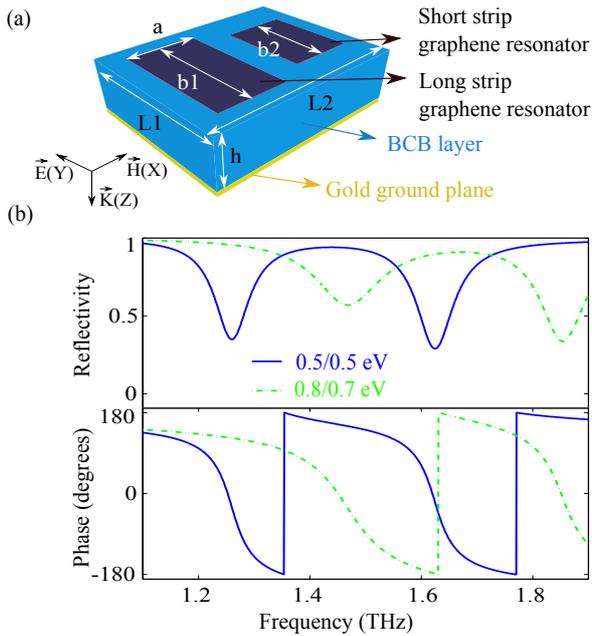}
\caption{(a) Schematics of the proposed TGLSR cell with dimensions of $h = 10\,\mu$m, $L1 = 28\,\mu$m, and $L2 = 40\,\mu$m. (b) Reflectivity and phase profile for the Fermi levels of 0.5/0.5 eV (blue line) and 0.8/0.7 eV (dot line) for long/short graphene strip resonators.}
\label{fig1} 
\end{figure}

The one atom thick graphene layer has shown strong metallic behavior at the THz regime. Its surface conductivity is mainly determined by intraband transition and is evaluated by~\cite{Surface-conductivity}
\begin{equation}\label{eq:1}
\sigma=-j\frac{e^2K_BT}{\pi\hbar^2(\omega-2j\Gamma)}(\frac{\mu_c}{K_BT}+2ln(e^{-\mu_c/K_BT}+1))
\end{equation}
where $e$ is the elementary charge, $T$ is the temperature, $\omega$ is the angular frequency, $\mu_c$ is the Fermi level, $\Gamma$ is a phenomenological scattering rate, $K_B$ is the Boltzmann's constant, and $\hbar$ is the reduced Planck's constant. The Fermi level can be changed by electrical gating or chemical doping, which gives great freedom to implement tunable conductivity. Several works on graphene based metasurfaces have been reported~\cite{APL-graphene-ribbon,Scientific-graphene-ribbon}. Previous works focused on TM excited plasmonic resonances in infinitely long graphene ribbons-due to the finite loss of graphene, only a maximum of 320$^{\circ}$ phase modulation was achieved~\cite{Scientific-graphene-ribbon}. In addition, the large phase modulation is only achieved in the resonance vicinity, which limits its working frequency range. The configuration of having infinitely long graphene ribbons cannot be extended to 2D wavefront control. To avoid those limitations, we propose to use two graphene long/short-strip resonators (TGLSR) as the building block. Two resonances are excited by having two different graphene long/short-strip resonators in one unit cell. Since the two resonances are determined by the Fermi level of each graphene strip, two independently controllable Fermi levels are available for tuning the metasurface, which extend the 2$\pi$ phase modulation to a large frequency range. 

Although the Fermi levels of graphene can be tuned by chemical doping, using gating voltage is the only practical way to change Fermi levels of graphene dynamically. Fermi levels ($E_f$) of graphene can be tuned by modifying the local carrier density - this is achieved, for example, by applying an external voltage~\cite{gating400V}
\begin{equation}\label{eq:2}
|E_f|=\hbar{V_f}\sqrt{\pi{N}}
\end{equation}
where graphene’s Fermi velocity is $V_f\approx1\times10^6m/s$, the total carrier density $N=\sqrt{n_0^2+\alpha^2|\Delta{V}|^2}$, $n_0$ is the residual carrier concentration induced by density fluctuations caused by charged impurities near the Dirac point and is varied across different graphene samples, $\alpha$ is the gate capacitance which is determined by the specific electrode configuration, $|\Delta{V}|=|V_{CNP}-V_g|$, with $V_{CNP}$ being the charge neutral point where numbers of electrons and holes are equal, $V_g$ is the applied external voltages. It is noted that $V_{CNP}$ is highly dependent on different graphene samples, e.g. $V_{CNP}=0.5 V$ in one graphene sample and it is moved to much higher value of $V_{CNP}=12 V$ after chemical doping~\cite{gating14V}. To determine the actual voltage $V_g$ for the corresponding Fermi levels, fitting parameters from experimental results are needed~\cite{gating400V}, and the reported values have been varied from several volts~\cite{Graphene_THz1,gating14V} to hundreds of volts~\cite{gating400V} which can be affected by the specific graphene samples, dielectric thickness and properties, and electrode configurations. In order to avoid complicating the underlying physics and not overwhelming readers, we have decided to use Fermi levels instead of voltages in the text, which is a normal practice in various studies~\cite{Weiren,APL-graphene-ribbon,Scientific-graphene-ribbon}.

Fig.~\ref{fig1}(a) shows the schematics of the TGLSR which consists of a 500 nm thick gold ground plane and two single layer graphene long/short-strips separated by a benzocyclobutene (BCB) dielectric layer. The two long/short-strips have the same width $a = 15\,\mu$m but different length $b_1 = 24\,\mu$m and $b_2 = 19\,\mu$m to have two resonances. Numerical simulations are performed by using the frequency domain solver in CST Microwave Studio. The global mesh is chosen as 16 cells per wavelength for the device and 6 for the vacuum background. The mesh is increased for the graphene resonators, with a maximum step width of 1 $\mu$m. Unit cell boundary is applied to both X and Y directions, and the Z direction has an open boundary. The relative permittivity of BCB dielectric is nearly constant from 0.5 to 5.4 THz~\cite{BCB_constant}, and we have used $\epsilon=2.67$ with a loss tangent of 0.012 (data obtained from~\cite{BCB}). Graphene can be modelled as either a conductance surface or a volumetric material with complex permittivity and the surface conductivity approach is typically more efficient. The dispersive complex permittivity of graphene is calculated from Eq.~\eqref{eq:1} and CST has a user defined materials icon which can be used to import complex permittivities from different materials for simulations. We have used a surface conductivity approach to model graphene with vanishing thickness in CST, which has a built-in dispersive graphene material calculated by using the complete Kubo formula~\cite{Surface-conductivity}. This approach is validated by reproducing the well-known results of graphene ribbons~\cite{Test-graphene-model}. When modeling graphene, the temperature of graphene is 300 K and the phenomenological scattering rate is assumed to be 0.11 meV~\cite{Surface-conductivity}. The material property of gold is taken from the CST built-in material library, which matches well with experimental results in the previous works~\cite{Liu2013,Liu2014}.

\begin{figure}[tb]
\centering
\protect\includegraphics[width=1.0\columnwidth]{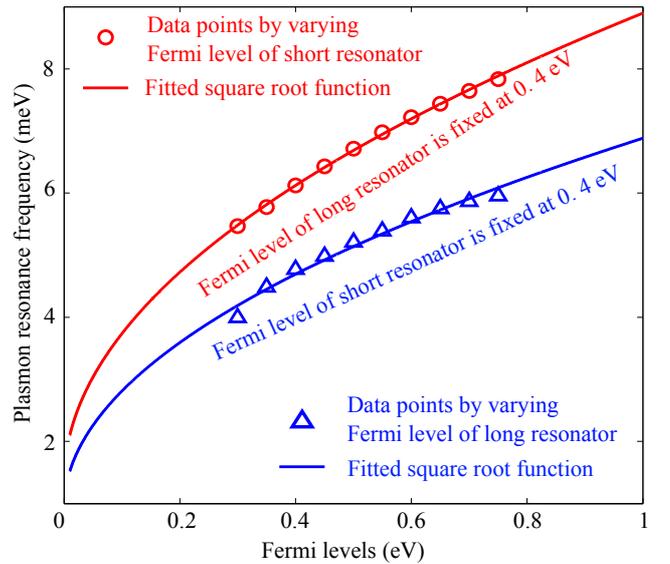}
\caption{\label{fig2} Square root functions are fitted separately from discrete data points by fixing one graphene resonator at 0.4 eV and varying  the other one from 0.3 to 0.75 eV. The red circles are the data points used to fit the square root function when fixing long-strip resonators at 0.4 eV and varying the Fermi level of short-strip resonator from 0.3 to 0.75 eV, which give the two fitting parameters $a=0.001347$ and $b=0.007549$. The blue triangles are the data points used to fit the square root function when fixing short-strip resonators at 0.4 eV and varying the Fermi level of long-strip resonator from 0.3 to 0.75 eV, which give the two fitting parameters $a=0.0009255$ and $b=0.005956$.}
\end{figure}

\begin{figure}[tb]
\centering
\protect\includegraphics[width=1\columnwidth]{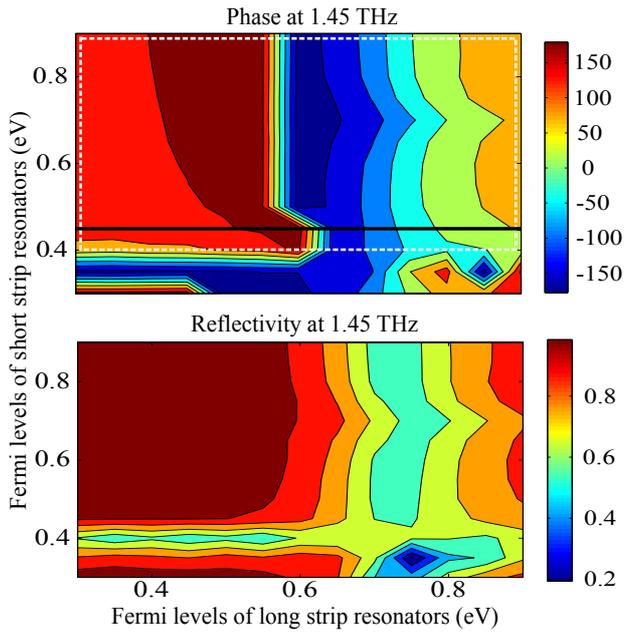}
\caption{\label{fig3} Reflection phase and amplitude for graphene resonators when Fermi levels of long/short graphene strips varying between 0.3 and 0.9 eV at 1.45 THz. The white square indicates the large region with full phase modulation. The black line stands for the covered phase values by varying Fermi levels of long resonators from 0.3 to 0.9 eV when the short resonators are fixed at 0.45 eV.}
\end{figure}

\begin{figure}[tb]
\centering
\protect\includegraphics[width=1.0\columnwidth]{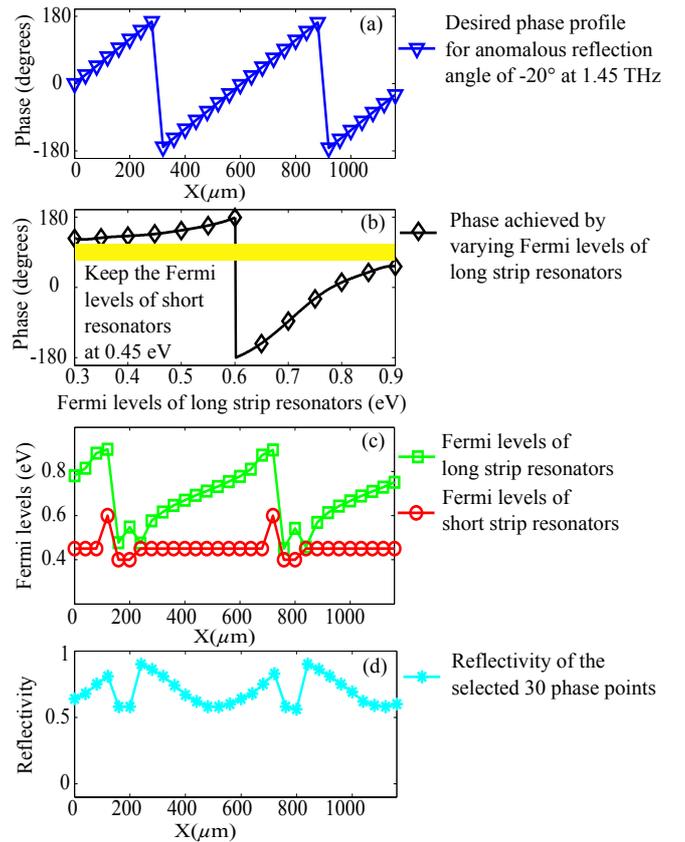}
\caption{\label{fig4} (a) Desired phase profile and the 30 phase points used to approximate this phase profile for an anomalous reflection angle of -20$^{\circ}$ at 1.45 THz. (b) The diamond line shows the phase values that can be covered by fixing the Fermi levels of short strip resonators at 0.45 eV while varying long resonators from 0.3 to 0.9 eV and the yellow region depicts the uncovered phase values. (c) The Fermi level combinations of the final selected long/short strip resonators to achieve the phase profile in Fig.~\ref{fig4}(a). (d) Reflection amplitude corresponding to those selected Fermi level combinations in Fig.~\ref{fig4}(c).}
\end{figure}

It is clear that two plasmon resonances are excited for an incident THz wave with electric field polarized along the Y direction as shown in the blue line of Fig.~\ref{fig1}(b). The resonance at 1.26 THz comes from the long strip resonator with a Fermi level of 0.5 eV and 1.63 THz is associated with the short strip resonator with a Fermi level of 0.5 eV. It is worthy noting that there is nearly a 2$\pi$ phase jump associated with each resonance, which means the full phase modulation can be achieved in a broad frequency range. More importantly, the two plasmon resonances are determined by their own Fermi levels, which effectively gives two independent variables with the capability of being controlled by applying external voltages. This is clearly confirmed by increasing the Fermi levels of long/short strip resonators from 0.5/0.5 eV to 0.8/0.7 eV, the two resonances shift from 1.26/1.63 to 1.47/1.85 THz accordingly.

To further quantitatively study the relationship between the plasmon resonances and Fermi levels, the plasmon resonances versus different Fermi levels are plotted in the discrete data points of Fig.~\ref{fig2}. The blue triangles show the shift of plasmon resonances when fixing the Fermi level of short-strip resonator at 0.4 eV and increasing the Fermi levels of long-strip resonator from 0.3 to 0.75 eV while the red circles show the long one is fixed at 0.4 eV while varying the short resonator from 0.3 to 0.75 eV, respectively. It is noted that the resonance frequency is converted to eV unit to better study its relation with Fermi levels. A damped oscillator can be used to study the plasmon resonance $\omega_p$ excited in graphene ribbons~\cite{Graphene_THz1}, and a power-law scaling with $\omega_p\propto|E_f|^{1/2}\propto{N}^{1/4}$ can quantitatively describe universal relation between $\omega_p$ and Fermi level $|E_f|$. Based on the data points, a square root function $\omega_p(|E_f|)=a+b|E_f|^{1/2}$, with two fitting parameters $a$ and $b$, is used to fit these data points and the fitted curves are shown in the solid lines in Fig.~\ref{fig2}. It is found that the two curves can rather accurately fit those data points with the corresponding square root functions. More accurate differential equations can be used to explain scaling behavior of plasmon resonance frequency, but the quantitative value of $\omega_p$ still cannot be solved analytically~\cite{Graphene_THz1}. Therefore, numerical simulations are mainly used to study THz response of the device.

To build a metasurface, it is necessary to have a continuous 2$\pi$ phase variations. Color map of phase and reflectivity at 1.45 THz is shown in Fig.~\ref{fig3} when varying Fermi levels of the two long/short-strip resonators from 0.3 to 0.9 eV, respectively. From the phase color map, it is clear that an exact 2$\pi$ phase modulation is not achievable by fixing the Fermi level of the short resonator in one value (e.g. 0.45 eV) and then varying the Fermi levels of the long resonator from 0.3 to 0.9 eV (shown as the black line in the phase color map). This result is consistent with the previous work which states that complete 2$\pi$ phase coverage is not reached due to the finite loss of graphene~\cite{Scientific-graphene-ribbon}. Thus by having two graphene resonators the problem is solved as we can select the uncovered phase values in other regions. The white square in Fig.~\ref{fig3} shows the large region of having full phase modulation.
To achieve similar results, previous design based on graphene ribbons have to change graphene ribbon width~\cite{APL-graphene-ribbon,Scientific-graphene-ribbon}, which is not practical after fabrication. With voltage controlled full phase modulation, dynamic reconfigurable metasurfaces are readily achievable based on the TGLSR.

\section{Terahertz Beam Steering}
\label{sec:2}

As a proof of concept, a THz beam steering is demonstrated based on the TGLSR. According to the generalized reflection law, the anomalous reflection angle for a normal incidence is determined by~\cite{Nanfang}
\begin{equation}
\label{eq:3}
\theta_r=arcsin(\frac{\lambda_0}{2\pi}\frac{d\Phi}{dx})
\end{equation}
where $\theta_r$ is the anomalous reflection angle, $\lambda_0$ is vacuum wavelength, and $\frac{d\Phi}{dx}$ is the phase discontinuity along
the interface. Various anomalous reflection angles are produced with different phase slopes of $\frac{d\Phi}{dx}$ as depicted in Eq.~\eqref{eq:3}. By applying different voltages to different graphene strip resonators, a dynamic beam steering function is implemented. Fig.~\ref{fig4}(a) shows the required phase profile for an anomalous reflection angle of -20$^{\circ}$ at 1.45 THz calculated from Eq.~\eqref{eq:3}. As it is not practical to have smooth phase variations, 30 discrete phase points (blue triangles) are used to approximate the phase profile. 
Moreover, there are some phase points that cannot be achieved (the yellow region in Fig.~\ref{fig4}(b)) by only varying Fermi levels of the long resonators when short resonators are fixed at 0.45 eV, as shown in Fig.~\ref{fig4}(b) which corresponds to the black line in Fig.~\ref{fig3}. Other Fermi level combinations are needed to fill the remaining phase points. Fig.~\ref{fig4}(c) shows the whole Fermi level combinations for long/short strip resonators to achieve the required phase profile in Fig.~\ref{fig4}(a).  
The reflectivity corresponding to those selected phase points are plotted in Fig.~\ref{fig4}(d), which shows a relative constant amplitude with a minimum reflectivity of 0.56.

\begin{figure}[tb]
\centering
\protect\includegraphics[width=1.0\columnwidth]{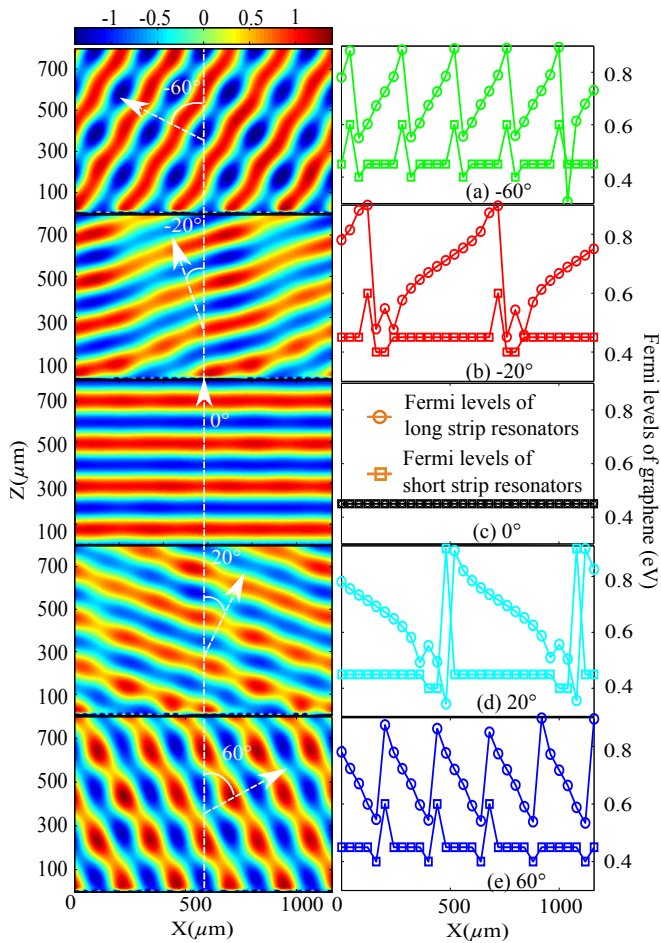}
\caption{\label{fig5} E-field of graphene metasurface based on the TGLSR for anomalous reflection angle of (a) -60$^{\circ}$, (b) -20$^{\circ}$, (c) 0$^{\circ}$, (d) 20$^{\circ}$ and (e) 60$^{\circ}$ at 1.45 THz with the incident field subtracted. The desired phase profiles for different anomalous reflection angles are calculated from ~\ref{eq:3}. The final selected Fermi level combinations for each case are denoted in the right part of Fig.~\ref{fig5}(a)-(e) accordingly.}
\end{figure}

Fig.~\ref{fig5} shows the simulated electric field distribution for different anomalous reflection angles with incident field subtracted at 1.45 THz. Anomalous reflection angles of -60$^{\circ}$, -20$^{\circ}$, 0$^{\circ}$, 20$^{\circ}$ and 60$^{\circ}$ in 1.45 THz are clearly observed through the E-field pattern in Fig.~\ref{fig5}(a), (b), (c), (d) and (e), respectively. A beam steering angle of as large as 120$^{\circ}$ has been demonstrated by having different Fermi levels of graphene resonators denoted in the right side of Fig.~\ref{fig5}(a)-(e) accordingly. It is noted that the Fermi levels of all the graphene resonators are fixed at 0.45 eV for 0$^{\circ}$ to confirm that no anomalous reflection is observed without the desired phase discontinuity. There are some distortions on the wavefront, which mainly come from numerical approximations and different reflection amplitude in each unit cell. 

\begin{figure}[tb]
\centering
\protect\includegraphics[width=1.0\columnwidth]{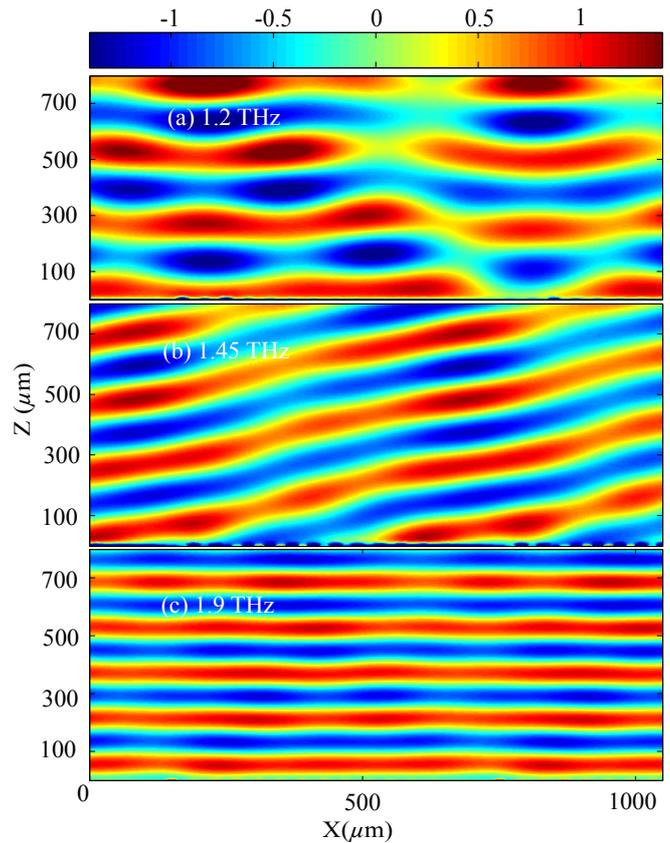}
\caption{\label{fig6} E-filed recorded at (a) 1.2 THz, (b) 1.45 THz and (c) 1.9 THz for TGLSR metasurface designed to have an anomalous reflection angle of -20$^{\circ}$ at 1.45 THz. It is clear that no anomalous reflection is achieved in other frequencies (1.2 and 1.9 THz) other than the designed working frequency (1.45 THz).}
\end{figure}

\begin{figure}[tb]
\centering
\protect\includegraphics[width=1.0\columnwidth]{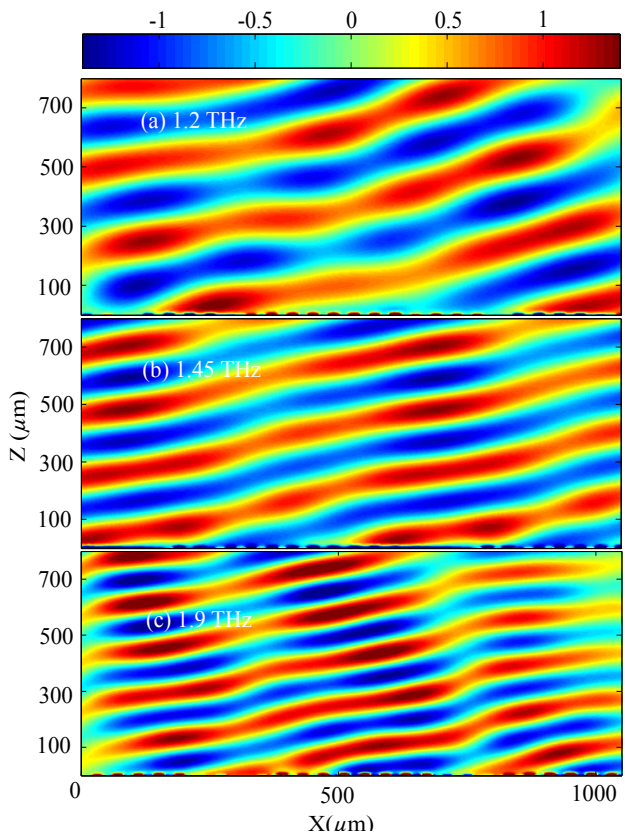}
\caption{\label{fig7} After dynamic phase modulation for each frequency, same anomalous reflection angle (-20$^{\circ}$) is achieved in (a) 1.2 THz, (b) 1.45 THz and (c) 1.9 THz, respectively. This confirms the TGLSR metasurface can achieve dynamic full phase modulation in large frequency range (700 GHz).}
\end{figure}

Since the discrete phase points in metasurfaces are produced by the corresponding resonances, a narrow working bandwidth is generally expected. This is confirmed by Fig.~\ref{fig6}(a) and (c) in which no clear anomalous reflection is observed at 1.2 and 1.9 THz for a metasurface designed to have an anomalous reflection angle of -20$^{\circ}$ at 1.45 THz. However there is a greater interest for a device of which working frequency can be tuned in real applications. Thus a dynamic 2$\pi$ phase modulation is needed to accomplish this. For the sake of simplicity, three different frequencies of 1.2, 1.45 and 1.9 THz are demonstrated to have the same anomalous reflection angle of -20$^{\circ}$, which are shown in Fig.~\ref{fig7}(a), (b) and (c) respectively. Through dynamic phase modulation by having different Fermi levels for different graphene resonators, the same beam steering angle is preserved for different working frequencies. 

\begin{figure}[tb]
\centering
\protect\includegraphics[width=1.0\columnwidth]{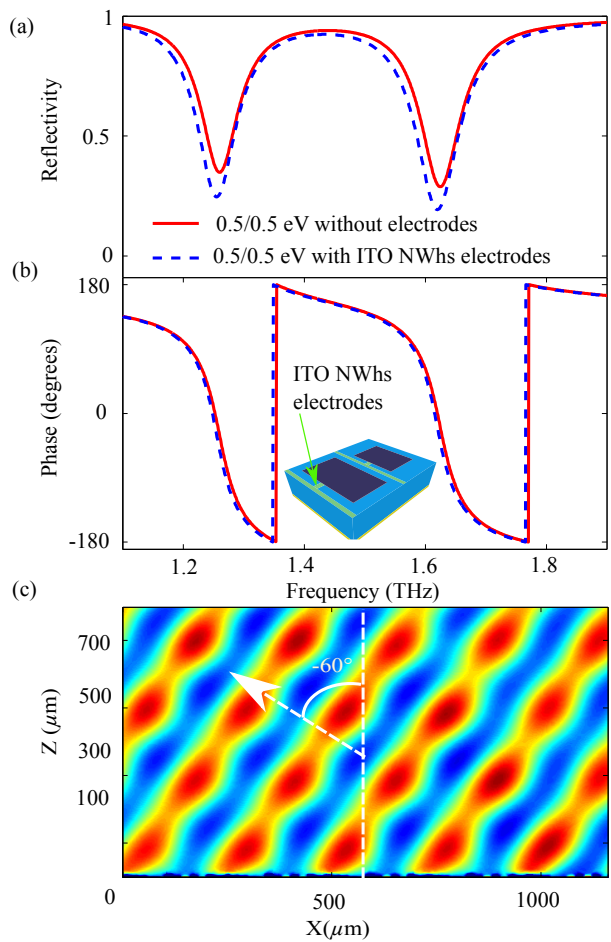}
\caption{\label{fig8} Effects of the ITO NWhs electrodes with 1 $\mu$m wide and 100 nm thick on the (a) reflectivity and (b) phase of the proposed TGLSR cell. (c) The re-simulated E-field of metasurface used in Fig.~\ref{fig5}(a) after adding the ITO NWhs electrodes, which shows that an anomalous reflection angles of -60$^{\circ}$ still can be achieved.}
\end{figure}

\begin{figure}[tb]
\centering
\protect\includegraphics[width=1.0\columnwidth]{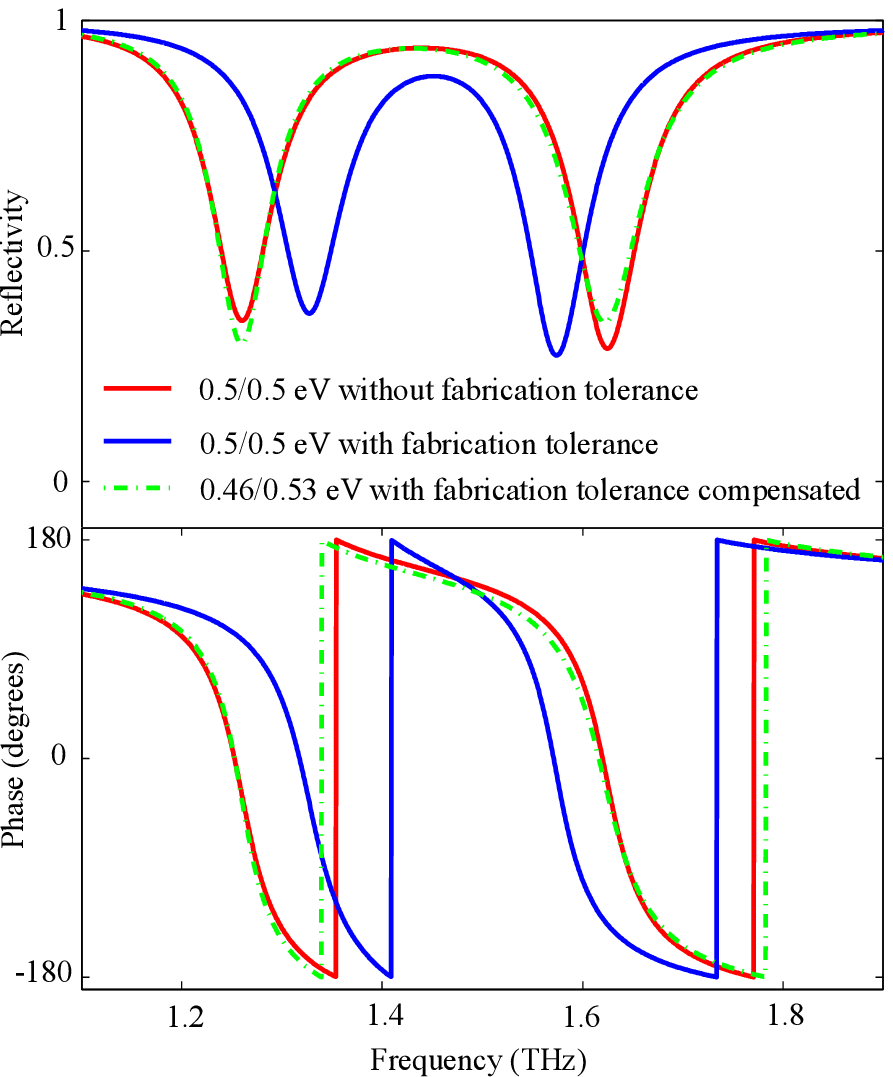}
\caption{\label{fig9} Effects of fabrication tolerance on the (a) reflectivity and (b) phase of the proposed TGLSR cell for Fermi levels of 0.5/0.5 eV. The 3\% fabrication tolerance can substantively shift the plasmon resonances. Through dynamic changing Fermi levels to 0.46/0.53 eV, the designed resonance can be restored as shown in the green dot line.}
\end{figure}

\section{Challenges of experimentally demonstrating this metasurface}
\label{sec:3}

Fabrication of graphene based tunable metamaterial devices has been demonstrated a few years ago: by using extraordinary optical transmission electrodes~\cite{gating400V} and ion-gel top gates~\cite{Graphene_THz1}, the carrier density has been largely changed in graphene. By having deep-subwavelength thin metallic array as electrodes, incident THz waves can be transmitted without being largely perturbed by these electrodes~\cite{gating400V}. An alternative and better approach to gate graphene is Indium-Tin-Oxide (ITO) Nanowhiskers (NWhs) which has been extensively studied as a transparent electrode in the THz range~\cite{ITO1,ITO2,ITO3}. Experimental works have shown that ITO NWhs exhibits transparency as high as 70\% up to 15 THz~\cite{ITO1} and has been reported to show transparency as high as 82\% in other works~\cite{ITO2}. Voltage controlled THz phase shifters based on transparent ITO NWhs have been demonstrated experimentally with 75\% transmission at THz frequencies~\cite{ITO2}. The highly transparency of ITO NWhs would better suit as electrodes for gating graphene without significantly affecting the incoming THz waves. We have placed ITO NWhs  electrodes with complex permittivity given in~\cite{ITO3}, and found that the whole response remains almost the same as without electrodes for Fermi levels of 0.5/0.5 eV, which is shown in the blue dot line of Fig.~\ref{fig8}(a) and (b). The metasurface used in Fig.~\ref{fig5}(a) is re-simulated by adding the ITO NWhs electrodes shown in the inset of Fig.~\ref{fig8}(b). The re-simulated E-field is shown in Fig.~\ref{fig8}(c) which further confirms that the ITO NWhs electrodes do not affect the optical performance of this device. Electron beam lithography (EBL) can be used to pattern and etch the ITO NWhs electrodes. Due to indium presence it is possible to make the NWhs by using an inherent micro-masking mechanism during the RIE process~\cite{Fouad}.

All the resonant metamaterial based devices need to address the problem of fabrication tolerance as its property is sensitive with geometry variations. We have applied a reasonable 3\% fabrication tolerance to the dimension of the  two graphene long/short-strip resonators and find the two resonance indeed shifts from 1.26/1.63 THz to 1.33/1.57 THz for Fermi levels of 0.5/0.5 eV as shown in Fig.~\ref{fig9}. By changing the Fermi levels of graphene to 0.46/0.53 eV, the two resonances come back to the designed values as shown in the green dot line in Fig.~\ref{fig9}. In a conventional metal based metasurface, the problem of fabrication tolerance cannot be easily solved. However for this graphene based metasurface, fabrication tolerances can be compensated by increasing or decreasing the Fermi levels of graphene controlled by external voltages. As it is shown in the green dot line, the designed response can be restored after compensating fabrication tolerance by changing graphene Fermi levels accordingly.

Through the existing experimental works on graphene based tunable metamaterial devices and additional simulations, we have shown that fabrication of this device is realistic, although challenging. Another important benefit of using this graphene metasurface is that fabrication tolerances can be accommodated by varying external voltages applied to graphene.

\section{Conclusions}
\label{sec:4}

In conclusion, we have proposed a novel graphene strip resonators as the fundamental cell for reconfigurable metasurfaces to have dynamic phase modulation in the THz frequencies. This configuration can be applied to any arbitrary wavefront manipulations with minimum of 56\%\ reflection efficiency. By independently controlling two plasmon resonances in graphene resonators, 2$\pi$ phase modulation can be accomplished in a large frequency range (700 GHz). Based on this metasurface, a THz beam steering function is demonstrated with steering angle of as large as 120$^{\circ}$. By rearranging different Fermi levels to different graphene resonators, the same beam steering angle is achieved from 1.2 to 1.9 THz. 
The most important advantage of this metasurface is that tunability can be implemented by applying external voltage to graphene resonators.
Gating graphene by the ITO Nanowhiskers electrodes which have high THz transparency do not affect THz performance of the device, which is confirmed through full-wave simulations. The fabrication tolerance can also be compensated through dynamically adjust the corresponding Fermi levels of graphene, which is a big advantage to the conventional metal based metasurfaces.
Moreover, this metasurface can be directly extended to 2D wavefront manipulation, which are highly beneficial for 2D flat lens, THz radar systems and communications.

\section*{}
\textbf{Acknowledgments}  This work is partially supported by the Asian Office of Aerospace Research and Development-U.S. Air Force (FA2386-15-1-4064) and the Australian Research Council. Liming Liu thanks the financial support from the China Scholarship Council (CSC).


\end{document}